\documentclass[aps,twocolumn,nofootinbib,preprintnumbers,superscriptaddress]{revtex4-1}
\pdfoutput=1

\usepackage{tabu}
\usepackage[usenames,dvipsnames,table]{xcolor}
\usepackage{graphicx,amsmath,amssymb,amsthm,multirow,array,bm,bbm,esint}
\usepackage[mathscr]{eucal}
\usepackage[bbgreekl]{mathbbol}
\usepackage{epsf,amsfonts}
\usepackage{slashed}
\usepackage{hyperref}

\usepackage{ulem}

\hypersetup{
    pdfstartview={FitH},    
    pdftitle={The eightfold way to dissipation},    
    pdfauthor={Felix Haehl, R.~Loganayagam, Mukund Rangamani},     
    colorlinks=true,       
    linkcolor=blue,          
    citecolor=red,        
    filecolor=magenta,      
    urlcolor=blue           
}
\definecolor{rust}{rgb}{0.8,0.2,0.2}

\newcommand{\beq}{\begin{equation}}
\newcommand{\eeq}{\end{equation}}
\newcommand{\bea}{\begin{eqnarray}}
\newcommand{\eea}{\end{eqnarray}}

\newcommand{\prn}[1]{\left ( #1 \right )}
\newcommand{\brk}[1]{\left [ #1 \right ]}

\newcommand{\half}{\frac{1}{2}}

\newcommand{\JH}{\mathrm{J}_{H}}

\newcommand{\THall}{\mathrm{T}_{H}}

\newcommand{\Lag}{{\mathcal L}}

\newcommand{\PS}{\text{H}_S}
\newcommand{\PV}{\text{H}_V}
\newcommand{\PF}{\text{H}_F}
\newcommand{\LS}{{\overline {\rm H}}_S}
\newcommand{\GV}{{\overline{\rm H}}_V}

\newcommand{\Ds}{\text{D}_s}
\newcommand{\Dv}{\text{D}_v}
\newcommand{\Diss}{\Delta}

\newcommand{\Kbeta}{{\bm{\beta}}}
\newcommand{\LambdaB}{\Lambda_{\bm{\beta}}}
\newcommand{\Bfields}{{\mathscr B}}

\newcommand{\hfields}{{\bm \Psi}}

\newcommand{\acc}{{\mathfrak a}}

\newcommand{\Eqfields}{{\mathscr K}}
\newcommand{\diffEq}{{\delta_{_\Eqfields}}}

\newcommand{\diffB}{\delta_{_{\Bfields}}}

\newcommand{\aheat}{{\mathfrak h}}
\newcommand{\acharge}{{\mathfrak n}}

\newcommand{\BerryG}{{\cal N}}
\newcommand{\BerryGA}{{\cal X}}
\newcommand{\BerryA}{{\cal S}}

\newcommand{\Wey}{{\scriptscriptstyle\mathcal{W}}}

\newcommand{\AWeyl}{\mathcal{W}}

\newcommand{\DWeyl}{\mathscr{D}^{\Wey}}
\newcommand{\RWeyl}{{}^\Wey R}

\newcommand{\gb}{\bar g}
\newcommand{\Ab}{\bar A}
\newcommand{\UT}{U(1)_{\scriptstyle{\sf T}}}
\newcommand{\AT}{{\sf A^{ \!{\scriptscriptstyle{(T)}}}}\!}
\newcommand{\LagT}{\Lag_{ \!{\scriptscriptstyle{T}}}}

\begin{document}

\title
{ The eightfold way to hydrodynamic dissipation
}
\preprint{DCPT-14/65}

\author{Felix M. Haehl}
\email{f.m.haehl@gmail.com}
\affiliation{Centre for Particle Theory \& Department of
Mathematical Sciences, Durham University, South Road, Durham DH1 3LE, United Kingdom}

\author{ R.\ Loganayagam}
\email{nayagam@gmail.com}
\affiliation{Institute for Advanced Study, Einstein Drive, Princeton, NJ 08540, USA.}

\author{Mukund Rangamani}
\email{mukund.rangamani@durham.ac.uk}
\affiliation{Centre for Particle Theory \& Department of
Mathematical Sciences, Durham University, South Road, Durham DH1 3LE, United Kingdom}

\begin{abstract}
We provide a complete characterization of hydrodynamic transport consistent with the second law of thermodynamics at arbitrary orders in the gradient expansion.  A key ingredient in facilitating this analysis is the notion of adiabatic hydrodynamics, which enables isolation of the genuinely dissipative parts of transport. We demonstrate that most transport is adiabatic. Furthermore, of the dissipative part, only terms at the leading order in gradient expansion are constrained to be sign-definite by the second law (as has been derived before).  

\end{abstract}


\maketitle

\section{Introduction}
\label{sec:intro}

Hydrodynamics is  the universal low energy  description  at sufficiently high temperatures of quantum systems  near thermal equilibrium. The dynamical fields are the  intensive parameters that describe the near thermal density matrix viz.,  temperature $T$, chemical potential $\mu$ along with
the fluid velocity  ($u^\mu$, $u^\mu\, u_\mu =-1$) which sets the local frame in which
the state appears thermal. The background sources are the metric $g_{\mu\nu}$ and the 
flavor sources $A_\mu$. The hydrodynamic state in a given background is then completely
characterized by a `thermal  vector' $\Kbeta^\mu$ and `thermal twist' $\LambdaB$ defined via 
\begin{equation}
\Bfields \equiv \bigg\{\Kbeta^\mu = \frac{u^\mu}{T}, \;\;\LambdaB = \frac{\mu}{T} -\Kbeta^\sigma \, A_\sigma \bigg\} \,.
\label{eq:hfields}
\end{equation}
The response to the background sources are encoded in the energy momentum tensor $(T^{\mu\nu})$ and charge current $(J^\mu)$   of the theory given in terms of  the hydrodynamic fields.
The dynamical equations are  the statements of conservation. In the presence of external sources and quantum anomalies (incorporated by the inflow Hall currents $\THall^{\mu\perp}$ and $\JH^\perp$) one has with $D_\mu = \nabla_\mu + [A_\mu, \cdot]$
\begin{equation}
\nabla_\nu T^{\mu\nu} =  J_\nu \cdot F^{\mu\nu}+\THall^{\mu\perp}\, \qquad
D_\nu J^\nu = \JH^\perp \,.
\label{eq:hydroCons}
\end{equation}

Phenomenologically, a hydrodynamicist finds {\em constitutive relations} that express the currents  in terms of the fields. The operators  are tensors built out of $\Bfields$, the background  sources $\{g_{\mu\nu}, A_\mu\}$, and their gradients, multiplied by  {\em transport coefficients} which are arbitrary scalar functions of $ T, \mu$.
 A-priori this `current algebra' formulation appears simple, since classifying such unrestricted tensors is a straightforward exercise in representation theory.

However, hydrodynamic currents should satisfy a further constraint \cite{landau} --  the second law of thermodynamics has to hold for arbitrary configurations of the low energy dynamics. In practice, one demands the existence of an entropy current $J_S^\mu$ with non-negative definite divergence $\nabla_\mu J^\mu_S \geq 0$.

At low orders in the gradient expansion it is not too hard to implement the constraints by hand and check what the second law implies; e.g.,  at one derivative order one finds viscosities and conductivities  need to be non-negative $\eta, \zeta, \sigma \geq 0$, which is physically intuitive. To date no complete classification has been obtained at higher orders, though the impressive analyses of \cite{Bhattacharyya:2012nq,Bhattacharyya:2013lha,Bhattacharyya:2014bha} come quite close. 

From a (Wilsonian) effective field theorist's perspective this phenomenological current algebra-like approach is unsatisfactory. Not only is the entropy current not associated with any underlying microscopic principle, but also the origin of dynamics as conservation  is obscure. 
A-priori a Wilsonian description for density matrices should involve working with doubled microscopic degrees of freedom, a la Schwinger-Keldysh or Martin-Siggia-Rose-Janssen-deDominicis. But one has yet to understand the  couplings between the two copies (influence functionals) allowed by the second law, which ought to encode information about dissipation (and curiously also anomalies \cite{Haehl:2013hoa}).

In this letter we describe a new framework for hydrodynamic effective field theories and provide a complete classification of transport. In particular,  hydrodynamic transport admits a natural decomposition into adiabatic and dissipative components: the latter contribute to entropy production, while the former don't. At low orders terms such as viscosities are dissipative;  a major surprise is that most higher order transport is adiabatic!

Adiabatic transport can be captured by an effective action with not only  Schwinger-Keldysh doubling of the sources, but also a new gauge principle,  $\UT$ {\em KMS gauge invariance}, with a gauge field $\AT$. This symmetry implies adiabaticity i.e., off-shell entropy conservation, providing thereby a rationale for $J^\mu_S$ (dissipative dynamics arises in the Higgs phase). We use this to prove an eightfold classification of adiabatic transport. Together with a  key theorem from \cite{Bhattacharyya:2013lha}, we further argue that dissipative hydrodynamic transport is constrained by the second law only at leading order in gradients.  In the following we will sketch the essential features of our construction;  details will appear in companion papers \cite{Haehl:2015pja}.

\section{Adiabatic hydrodynamics and the Eightfold Way}
\label{sec:adiabatic}

The key ingredient of our analysis which enables the classification scheme is the notion of adiabaticity. The main complications in hydrodynamics arise from attempting to implement the second law of thermodynamics on-shell. Significant simplification can be achieved by taking the constraints off-shell. One natural way to do this is to extend the inequality $\nabla_\mu J^\mu_S \geq 0$ to an off-shell statement by the addition of the dynamical equations of motion with Lagrange multipliers \cite{Liu:1972nr}. Choosing the Lagrange multipliers for the energy-momentum and charge conservation to be the  hydrodynamic fields  implies that
\begin{equation}\label{eq:Adiabaticity}
\begin{split}
\nabla_\mu J_S^\mu &+ \Kbeta_\mu\prn{\nabla_\nu T^{\mu\nu}-J_\nu \cdot F^{\mu\nu}-\THall^{\mu\perp}}\\
&+ (\LambdaB+\Kbeta^\lambda A_\lambda) \cdot \prn{D_\nu J^\nu-\JH^\perp} =\Diss \geq0
\,,
\end{split}
\end{equation}
with $\Diss$ capturing the dissipation and ``$\cdot$" denotes flavour index contraction.

While taking the second-law inequality off-shell allows us to ignore on-shell dynamics, one can obtain the most stringent conditions by examining the boundary of the domain where we marginally satisfy the constraint. We define an adiabatic fluid as one where the off-shell entropy production is compensated for precisely by energy-momentum and charge transport. We thus motivate the study of the {\em adiabaticity equation} obtained from \eqref{eq:Adiabaticity} by setting $\Diss =0$. We will refer to the set of functionals 
$\{J_S^\mu,T^{\mu\nu},J^\mu\}$ that satisfy $\Diss =0$ as  the adiabatic constitutive relations.  

Implications of adiabaticity were first studied in the context of anomalous transport in \cite{Loganayagam:2011mu} and are explored in greater detail in \cite{Haehl:2015pja}. In the following we will quote some of the salient results of our analysis and explain how it  helps with the taxonomy.

Intuitively, the notion of adiabaticity is an off-shell generalization of non-dissipativeness; imposing \eqref{eq:hydroCons} we learn that the entropy current has to be conserved on-shell.  Moreover, apart from quantum anomalies encoded by the Hall currents, the contributions at each order in the gradient expansion can be decoupled. It is quite remarkable  that this corner of the hydrodynamic constitutive relations is sufficient to delineate all the constraints on transport. We will first outline different classes of solutions to the adiabaticity equation \eqref{eq:Adiabaticity} and then in \S\ref{sec:algo} explain how it can be utilized for taxonomic purposes.

\begin{figure}[htbp]
\begin{center}
\includegraphics[width=3in]{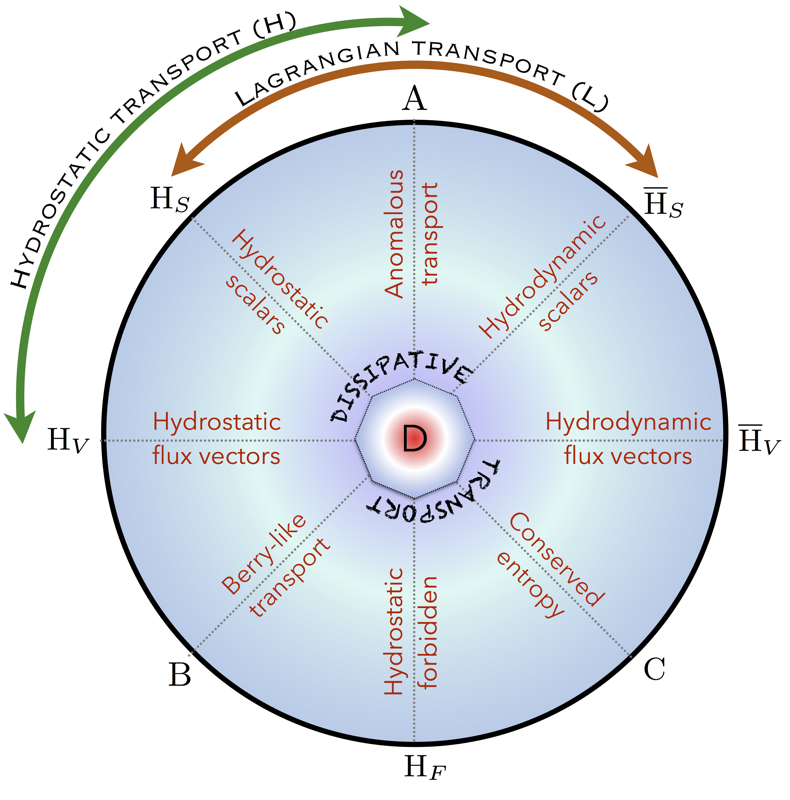}
\begin{picture}(0,0)
\setlength{\unitlength}{1cm}
\end{picture}
\caption{The eightfold way of hydrodynamic transport. }
\label{fig:eightfold}
\end{center}
\end{figure}

The adiabatic transport finds a natural classification into eight primary classes -- see Fig.~\ref{fig:eightfold}. We emphasize 
that adiabatic constitutive relations encode those transport coefficients which never appear in the expression for entropy production.
Together with Class D (dissipative) we exhaust all transport along this eightfold path.

To understand the nomenclature and taxonomy, let us start with Class A which comprises of transport fixed by the  quantum anomalies of the QFT. Such anomalous transport gives a particular solution to the adiabaticity equation 
\eqref{eq:Adiabaticity}, cf., \cite{Loganayagam:2011mu} -- the anomalous Hall currents can be viewed as inhomogeneous source terms. This allows us to dispense with them once and for all and focus thence on the non-anomalous adiabaticity equation.

The simplest  solutions to \eqref{eq:Adiabaticity} can be obtained by restricting to {\em hydrostatic equilibrium} (Class H). One subjects the fluid  to arbitrary slowly varying, time-independent external sources $\{g_{\mu\nu}, A_\mu\}$. The background time-independence implies the existence of  Killing vector and gauge  transformation, $\Eqfields= \{K^\mu,\Lambda_K\}$, with  $\diffEq g_{\mu\nu} = \diffEq A_\mu =0$. Identifying the hydrodynamic fields with these background isometries  $\Kbeta^\mu = K^\mu$, $\LambdaB = \Lambda_K$ solves \eqref{eq:Adiabaticity}. This information can equivalently be encoded in a hydrostatic partition function \cite{Banerjee:2012iz,Jensen:2012jh} which is the generating functional of (Euclidean) current correlators. Varying this partition function, we can then obtain a class of constitutive relations that solve \eqref{eq:Adiabaticity}.

The partition function has two distinct components: hydrostatic scalars $\PS$ and  vectors $\PV$. The transformation properties refer to the transverse spatial manifold obtained by reducing along the (timelike) isometry direction.\footnote{These scalars and vectors in the generating function generate the tensor operators for the currents (including part of dissipative transport) upon variation with respect to the background metric and gauge field.} The scalars $\PS$ are terms one is most familiar with; e.g., the pressure $p$ as a functional of intensive parameters (which now are determined by the background Killing fields). The vectors $P^\sigma$ in $\PV$ are  both transverse to the Killing field and conserved on the co-dimension one achronal slice, i.e.,  $K_\sigma\,P^\sigma = \nabla_\sigma P^\sigma =0 $.

Hydrostatics fixes a part of the constitutive relations by imposing relations between a-priori independent transport coefficients   \cite{Banerjee:2012iz}. These relations (Class $\PF$)  capture the fact that non-vanishing hydrostatic currents expressed as independent tensor structures in equilibrium, arise from a single partition function. More importantly, dangerous terms which can produce sign-indefinite divergence of entropy current are eliminated in Class $\PF$.

The second set of solutions of \eqref{eq:Adiabaticity} are generated by generalizing the scalar part of the partition function to time-dependent configurations, a la Landau-Ginzburg. We call these {\em Lagrangian} (Class L) solutions, since one can find a local Lagrangian (or Landau-Ginzburg free-energy) of the hydrodynamic fields and sources $\Lag\brk{\Kbeta^\mu, \LambdaB, g_{\mu\nu}, A_\mu}$. The currents are defined through standard variational calculus which can be expressed after suitable integrations by parts as
\begin{align}\label{eq:LagVar}
\frac{1}{\sqrt{-g}}\delta&\prn{\sqrt{-g}\ \Lag} =  \half \; T^{\mu\nu}\;\delta g_{\mu\nu} + J^\mu \cdot \delta A_\mu +T \,\aheat_\sigma \;\delta \Kbeta^\sigma
\nonumber \\&
+  T\, \acharge \cdot \prn{\delta\LambdaB+ A_\sigma \,\delta \Kbeta^\sigma}
+ \text{bdy. terms}
\end{align}
while the entropy density is defined as (nb: $J_S^\mu = s\, u^\mu$)
\begin{align*}
s  \equiv
\prn{\frac{1}{\sqrt{-g}}\; \frac{\delta}{\delta T}\; \int \sqrt{-g}\; \Lag\brk{\hfields} \; }\bigg|_{\{u^\sigma, \,\mu,
\,g_{\alpha\beta},\,A_\alpha\} = \text{fixed}}
\end{align*}
with $\hfields \equiv \{\Kbeta^\mu, \LambdaB, g_{\mu\nu}, A_\mu\}$.
Diffeomorphism and gauge invariance of $\Lag$ together imply a set of Bianchi identities, which together with the definition of $J^\mu_S$ suffices to show that \eqref{eq:Adiabaticity} is satisfied.\footnote{Obtaining the dynamical equations of motion i.e., conservation in Class L requires a constrained variational principle wherein one only considers variations in the Lie orbit of a reference configuration, cf., 
\cite{Haehl:2015pja}. Class L  is equivalent up to a Legendre transform to the non-dissipative effective action formalism developed in 
\cite{Dubovsky:2011sj,Bhattacharya:2012zx}. Thus the effective action describes a proper subset of adiabatic constitutive relations.} In the above equation, one can interpret $\{\aheat_\sigma,\acharge\}$ as characterizing the adiabatic heat current
and adiabatic charge density which satisfy a relation of the form $Ts+\mu\cdot \acharge = -u^\sigma \aheat_\sigma$.

It is intuitively clear that by restricting Class L solutions to hydrostatics we recover the partition function scalars $\PS$. As a result one can write $\text{L} = \PS \cup \LS$ with $\LS$ denoting scalar invariants that vanish identically in hydrostatics; hence hydrostatic scalars take values in a coset manifold
$\text{L}/\LS$.

There are two other adiabatic constitutive relations which are non-hydrostatic but non-dissipative. One class of adiabatic constitutive relations describe  Berry-like transport (Class B) 
which can be parameterized as
\begin{align}\label{eq:introClassB}
(T^{\mu\nu})_B
&\equiv -\frac{1}{2}{\cal N}^{(\mu\nu)(\alpha\beta)}\diffB  g_{\alpha\beta} + {\cal X}^{(\mu \nu) \alpha} \cdot \diffB  A_\alpha
\nonumber \\
(J^\alpha)_B
&\equiv  -\frac{1}{2}\, {\cal X}^{(\mu \nu) \alpha} \diffB  g_{\mu\nu}- {\cal S}^{[\alpha\beta]} \cdot \diffB  A_\beta \,.
\end{align}
Here ${\cal N}^{(\mu\nu)(\alpha\beta)}= -{\cal N}^{(\alpha\beta)(\mu\nu)}$, 
${\cal X}^{\mu \nu \alpha}$, and ${\cal S}^{\alpha\beta}$  are arbitrary local functionals of $\hfields$ with indicated (anti)symmetry 
properties, such that along with  
$J^\mu_S = - \Kbeta_\nu \,T^{\mu\nu} - \, \frac{\mu}{T}\, J^\mu$, the adiabaticity equation is satisfied \cite{Haehl:2015pja}.\footnote{Here $\diffB$ denotes Lie derivatives implementing diffeomorphisms and flavour gauge transformations by $\Bfields$, i.e., $\diffB g_{\mu\nu} =2\, \nabla_{(\mu}\Kbeta_{\nu)}$ and
 $\diffB A_\mu = D_\mu(\LambdaB+\Kbeta^\sigma\,A_\sigma) + \Kbeta^\nu\, F_{\nu\mu}$.} A prime example for structures of the type (\ref{eq:introClassB}) are the parity odd shear tensor in $3$ dimensions which contributes to Hall viscosity (Class B). Thus, the tensors ${\cal N}^{\mu\nu\alpha\beta},{\cal X}^{\mu \nu \alpha}$, and ${\cal S}^{\alpha\beta}$
can be thought of as a generalization of the notion of odd viscosities and conductivities.

We will denote the other class as Class $\GV$ which can be  parameterized as:
\begin{equation}\label{eq:TJVec}
\begin{split}
(T^{\mu\nu})_{\GV} &\equiv
	  \half \brk{D_\rho{\mathfrak C}_{\BerryG}^{\rho (\mu\nu)(\alpha\beta)}
	  \, \diffB  g_{\alpha\beta} + 2\ {\mathfrak C}_{\BerryG}^{\rho (\mu\nu)(\alpha\beta)}
	  \, D_\rho \diffB  g_{\alpha\beta}} \\
&\qquad 	+ D_\rho{\mathfrak C}_{\BerryGA}^{\rho(\mu \nu) \alpha} \cdot \diffB  A_\alpha
	   + 2\ {\mathfrak C}_{\BerryGA}^{\rho(\mu \nu) \alpha} \cdot
  	  \, D_\rho \diffB  A_\alpha \\
(J^\alpha)_{\GV} &\equiv
	 \half \brk{ D_\rho{\mathfrak C}_{\BerryGA}^{\rho(\mu \nu) \alpha}  \diffB  g_{\mu\nu}
	   + 2\ {\mathfrak C}_{\BerryGA}^{\rho(\mu \nu) \alpha}
	  \, D_\rho \diffB  g_{\mu\nu} } \\
&\qquad 	+ D_\rho{\mathfrak C}_{\BerryA}^{\rho(\alpha\beta)} \cdot \diffB  A_\beta
	   + 2\ {\mathfrak C}_{\BerryA}^{\rho (\alpha\beta)} \cdot
  	  \, D_\rho \diffB  A_\beta
\end{split}
\end{equation}
where $ {\mathfrak C}_{\BerryG}^{\rho (\mu\nu)(\alpha\beta)}= {\mathfrak C}_{\BerryG}^{\rho (\alpha\beta)(\mu\nu)}$. The entropy current
has a similar form as in Class B along with an additional  contribution which is quadratic in $\diffB g_{\mu\nu}$ and $\diffB A_\mu$. Finally, we 
have exactly conserved vectors (Class C) which can be added to the entropy current without modification of the  constitutive relations. They
describe possible topological states which transport entropy but no charge or energy.  We claim that the above classification is exhaustive:

\medskip
\noindent
{\bf Theorem:} The eightfold classes of adiabatic hydrodynamic transport can be obtained from a scalar Lagrangian density $\LagT\brk{\Kbeta^\mu, \LambdaB, g_{\mu\nu}, A_{\mu}, 
\gb_{\mu\nu},\Ab_{\mu}, \AT_\mu}$:
\begin{align}
\LagT &=\frac{1}{2}\, T^{\mu\nu}\, \gb_{\mu\nu} + J^\mu\cdot \Ab_{\mu} 
\nonumber \\
& \quad + \;\left(J_S^\sigma+\Kbeta_\nu T^{\nu\sigma}+(\LambdaB+\Kbeta^\nu A_\nu)\cdot J^\sigma\right)\AT_\sigma \,.
\label{}
\end{align}
 As indicated the Lagrangian density depends not only on the hydrodynamic fields and the background sources, but also the `Schwinger-Keldysh' partners of the sources $\{\gb_{\mu\nu}, \Ab_\mu\}$ and a new KMS gauge field $\AT_\mu$. This Lagrangian is invariant under diffeomorphisms and gauge transformations\footnote{Anomalies if present are dealt with using the inflow mechanism \cite{Callan:1984sa}. $\LagT$  then includes a topological theory  in $d+1$ dimensions coupled to the physical $d$-dimensional QFT (at the boundary/edge).} and under $\UT$ which acts only on the sources as a diffeomorphism or gauge transformation along $\Bfields$. The $\UT$ gauge invariance implies a Bianchi identity, which is nothing but the adiabaticity equation \eqref{eq:Adiabaticity}. Furthermore, a constrained variational principle for the fields $\{\Kbeta^\mu,\LambdaB\}$ ensures that the dynamics of the theory is simply given by conservation. We anticipate that the KMS gauge field plays a crucial role in implementing non-equilibrium fluctuation-dissipation relations which follow from the KMS condition; its significance both in hydrodynamic effective field theories as well as in holography will be discussed in a future work
 \cite{Haehl:2015fk}.

\section{The route to dissipation}
\label{sec:algo}

Having classified solutions to the adiabaticity equation let us now turn to the characterization of hydrodynamic transport including dissipative terms (Class D). We will do so by first systematically eliminating all of the adiabatic transport  by the following algorithm:
\begin{enumerate}
\item Enumerate the total number of transport coefficients, $\text{Tot}_{k\partial}$, at the $k^{\rm th}$ order in the derivative expansion. This can be done by either working in a preferred fluid frame, or more generally  by classifying frame-invariant scalar, vector and tensor data.
\item Find the particular solution to the anomaly induced transport (if any); this fixes all terms in Class A.
\item Restrict to hydrostatic equilibrium. The (independent) non-vanishing scalar fields and transverse conserved vectors determine $\PS$ and $\PV$ respectively (after factoring out terms which are related up to total derivatives), which parameterize the (Euclidean) partition function \cite{Banerjee:2012iz,Jensen:2012jh}.
\item Classify the number of tensor structures entering constitutive relations that survive the hydrostatic limit. Since they are to be determined from $\PS$ and $\PV$ respectively, we should have a number of hydrostatic relations $\PF$. In general the hydrostatic constrained transport coefficients are given as  linear differential combinations of unconstrained ones. 
\item Determine the Class L scalars that vanish in hydrostatic equilibrium $\LS$ from the list of frame invariant scalars after throwing out terms in $\PS$ (and those related by total derivatives). 
\item Find all solutions to Class B  and $\GV$ terms at the desired order in the gradient expansion by classifying potential tensor structures $\{{\cal N}, {\cal X},{\cal S}\}$
 and $\{{\mathfrak C}_{\BerryG},{\mathfrak C}_{\BerryGA},{\mathfrak C}_{\BerryA}\}$ respectively. We have now solved for the adiabatic part of hydrodynamics.
\item The remainder of transport is dissipative and contributes to $\Delta \neq 0$.  Class D is subdivided into two classes: terms constrained by the second law lie in Class $\Dv$, while those in
Class $\Ds$  contribute sub-dominantly to entropy production and are arbitrary. The goal at this stage is to isolate the $\Dv$ terms; fortunately they only show up only at the leading order in the gradient expansion ($k=1$); for $k\geq 1$ all dissipative terms are in Class $\Ds$ (cf., \cite{Bhattacharyya:2013lha,Bhattacharyya:2014bha}). 
\item Finally, Class $\Ds$ can be written in terms of dissipative tensor structures using the same formalism employed for Class B, except now we pick a different symmetry structure to  ensure $\Diss \neq 0$. 
\end{enumerate}
Steps 1-6  can be implemented straightforwardly in the $\UT$ invariant  $\LagT$, but we will exemplify this algorithm by a more pedestrian approach below. In Table \ref{tab:hclassify} we provide a classification of  transport for few hydrodynamic systems up to second order in gradient expansion.

\begin{table}[h]
\begin{tabular}{||c||c|c|c|c|c|c|c|c|c|c|}
\hline
Fluid Type&  Tot & $\PS$& $\LS$& $\PF$ & $\PV$ & A & B & $\GV$  & $D$
 \\
   \hline
Neutral  $1\partial$&  2 & 0 & 0 & 0 & 0 & 0 & 0  & 0 & 2 \\
Neutral $2\partial$ &  15 & 3 & 2 & 5 & 0 & 0 & 2  & 0  &3 \\
Weyl neutral $2\partial$ &  5 & 2 & 1 & 0 & 0 & 0 & 1  & 0  &1 \\
\hline
Charged $1\partial$ & 5&  0 & 0 & 2 & 0 & 0 & 0  & 0  &3 \\
Charged $2\partial$ & 51 &7 &5 & 17 & 0 & 0 & 11 & 2  &9 \\
\hline
\end{tabular}
\caption{Transport taxonomy for some simple (parity-even) fluid systems in $d\geq 4$. The fluid type refers to whether we describe pure energy-momentum transport (neutral) or transport with a single global symmetry (charged). We have indicated the derivative order at which we are working by $k\partial$. }
\label{tab:hclassify}
\end{table}

\section{An example: Weyl invariant neutral fluid}
\label{sec:weylinv}

To illustrate our construction consider a (parity-even) Weyl invariant neutral fluid which has been studied extensively in the holographic context \cite{Baier:2007ix,Bhattacharyya:2008jc,Bhattacharyya:2008mz}.  Weyl invariance implies that the stress tensor must be traceless and built out of Weyl covariant tensors.  Our classification suggests the following constitutive relation written in a basis adapted to the eightfold way:\footnote{The fluid tensors are defined via the decomposition $\nabla_\mu\,u_\nu = \sigma_{(\mu\nu)} + \omega_{[\mu\nu]} + \frac{1}{d-1}\,
\Theta\, (g_{\mu\nu}+ u_\mu\,u_\nu) - \acc_\nu\,u_\mu$ and $^{<>}$ denotes the symmetric, transverse (to $u^\mu$) traceless projection. The Weyl covariant derivative \cite{Loganayagam:2008is} (and associated curvatures) preserve homogeneity under conformal rescaling. In particular, $\RWeyl =R + 2(d-1)\left(\nabla_\alpha \AWeyl^\alpha - \frac{d-2}{2}\, \AWeyl^2\right)$, with Weyl connection $\AWeyl_\mu= \acc_\mu -\frac{\Theta}{d-1}\, u_\mu$ appears in Eq.~\eqref{eq:weyl2lambdaFG}.}
\begin{align}
T^{\mu\nu} &= p\left( d\, u^\mu u^\nu + g^{\mu\nu} \right) -2\, \eta\, \sigma^{\mu\nu} 
\nonumber \\
&+ (\lambda_1 - \kappa)\, \sigma^{<\mu\alpha} \sigma_{\alpha}^{\nu>} + \left(\lambda_2 + 2\,\tau -2 \kappa\right)\, \sigma^{<\mu\alpha} \omega_{\alpha}^{\ \nu>} 
\nonumber \\
& 
+ \tau \left( u^\alpha \DWeyl_\alpha \sigma^{\mu\nu} - 2\, \sigma^{<\mu\alpha} \omega_{\alpha}^{\ \nu>}\right)  
+ \lambda_3 \, \omega^{<\mu\alpha} \omega_{\alpha}^{\ \nu>}
\nonumber \\
&+ \kappa \,\left(C^{\mu\alpha\nu\beta}\,u_\alpha\,u_\beta + \sigma^{<\mu\alpha} \sigma_{\alpha}^{\nu>} + 2\, \sigma^{<\mu\alpha} \omega_{\alpha}^{\ \nu>} \right) .
\label{}
\end{align}

To obtain this note that for a neutral fluid there are no anomalies so $A=0$. At first order there is only a Class D term $\eta\, \sigma_{\mu\nu}$ which contributes to $\Diss = 2\,\eta\, \sigma^2$, leading to $\eta \geq 0$ (shear viscosity is non-negative). At second order we have two hydrostatic scalars $\omega_{\mu\nu}\, \omega^{\nu\mu}$ and $\RWeyl$; hence $\PS=2$ corresponding to $\lambda_3$ and $\kappa$ terms. As $\sigma_{\mu\nu}$ vanishes in hydrostatics only two tensors survive the limit; thus there are no constraints, $\PF =0$. There are no transverse vectors and so $\PV =\GV =0$. Surprisingly $ \left(\lambda_2 + 2\,\tau -2 \kappa\right)\ \sigma^{<\mu\alpha} \omega_{\alpha}^{\ \nu>}$ is a Class B term  -- it  can be obtained from ${\cal N}^{[(\mu\nu)|(\alpha\beta)]} \sim
(\lambda_2+2\,\tau -2\,\kappa)\, ( \omega^{\mu\alpha} P^{\nu \beta}   + \text{perms}. )$. There is  one non-hydrostatic scalar $\sigma^2$ which is in $\LS$
corresponding to $\tau$ term above. This leaves us with one Class D term which can be inferred to be $(\lambda_1 - \kappa)\sigma^{<\mu\alpha} \sigma_{\alpha}^{\ \nu>}$. Its contribution to entropy production is 
$\nabla_\mu J_S^\mu  \sim (\lambda_1 -\kappa) \,\sigma_{\alpha\nu} \sigma^{\nu\beta}\sigma^\alpha_{\ \beta}$. This being sub-dominant to the leading order $\eta\, \sigma^2$ entropy production, it follows that $(\lambda_1 - \kappa)$ belongs to Class $\Ds$. 

While this completes the classification, we note one rather intriguing fact. For holographic fluids dual to two derivative gravity, the  second order constitutive relations (cf., \cite{Bhattacharyya:2008mz}) can be derived from a Class L Lagrangian:
\begin{align}
\Lag^\Wey &=-\frac{1}{16\pi G_{d+1}}
\prn{\frac{4\pi T}{d}}^{d-2}\nonumber \\
&\quad \times \brk{ \frac{\RWeyl}{(d-2)}  +\half  \, \omega^2 +\frac{1}{\,d}\,  \text{Har}\prn{\frac{2}{d}-1}\, \sigma^2}
\label{eq:weyl2lambdaFG}
\end{align}
where $\text{Har}(x) = \gamma_e + \frac{\Gamma'(x)}{\Gamma(x)}$ is the Harmonic number function
($\gamma_e$ is Euler's constant). The first two terms are in $\PS$ while $\sigma^2 \in \LS$ and they give contributions to each of the five second order transport coefficients. We therefore have two relations:
\begin{equation}
\lambda_2 + 2\,\tau -2 \kappa=0 \,,\qquad  \lambda_1-\kappa=0\,.
\label{eq:weylrel}
\end{equation}
Eliminating $\kappa$ we have $\tau = \lambda_1 - \frac{1}{2}\, \lambda_2$ which was argued to in fact be a universal property of two derivative gravity theories  \cite{Haack:2008xx}.
Curiously, the first relation is also obeyed in kinetic theory to the orders in which computations are available \cite{York:2008rr}. 
We advance this as the evidence that our eightfold classification explains various hitherto unexplained coincidences in both perturbative transport calculations and non-perturbative results from AdS/CFT.

The second relation in \eqref{eq:weylrel} suggests that the sub-leading entropy production from $(\lambda_1-\kappa)$ is  absent in AdS black holes.\footnote{The relation between $\{\tau, \kappa,\lambda_2\}$ appears not to hold in higher derivative gravitation theories. It however must be borne in mind that generic higher derivative gravity theories are unlikely to be dual to unitary QFTs. We thank 
S. Grozdanov, E. Shaverin, A. Starinets and A.Yarom for sharing their results.} Inspired by earlier observations  regarding lower bound of shear viscosity $\eta/s \geq \frac{1}{4\pi}$ \cite{Kovtun:2004de}, we conjecture that holographic fluids obtained in the long-wavelength limit of strongly interacting quantum systems obey a {\em principle of minimal dissipation}. The fluid/gravity correspondence provides the shortest path in the eightfold way: AdS black holes scramble fast to thermalize, but are slow to dissipate!

\begin{acknowledgements}

It is a pleasure to thank K.~Balasubramanian, J.~Bhattacharya, S.~Bhattacharyya, V.~Hubeny, K.~Jensen, H.~Liu, J.~Maldacena, G.~Moore, S.~Minwalla, D.T.~Son,
A.~Starinets and A.~Yarom for enjoyable discussions on various aspects of hydrodynamics. 

FH  and MR would like to thank the IAS, Princeton for hospitality during the course of this project. MR would in addition like to thank YITP, Kyoto, U. Amsterdam and Aspen Center for  Physics for hospitality during the course of this project. FH is supported by a Durham Doctoral Fellowship. 
RL is supported by Institute for Advanced Study, Princeton. MR was supported in part by the Ambrose Monell foundation,  by the STFC Consolidated Grants ST/J000426/1 and  ST/L000407/1, by the NSF grant under Grant No. PHY-1066293, and by the ERC Consolidator Grant Agreement ERC-2013-CoG-615443: SPiN.

\end{acknowledgements}

\appendix



\newpage
\begin{thebibliography}{21}
\expandafter\ifx\csname natexlab\endcsname\relax\def\natexlab#1{#1}\fi
\expandafter\ifx\csname bibnamefont\endcsname\relax
  \def\bibnamefont#1{#1}\fi
\expandafter\ifx\csname bibfnamefont\endcsname\relax
  \def\bibfnamefont#1{#1}\fi
\expandafter\ifx\csname citenamefont\endcsname\relax
  \def\citenamefont#1{#1}\fi
\expandafter\ifx\csname url\endcsname\relax
  \def\url#1{\texttt{#1}}\fi
\expandafter\ifx\csname urlprefix\endcsname\relax\def\urlprefix{URL }\fi
\providecommand{\bibinfo}[2]{#2}
\providecommand{\eprint}[2][]{\url{#2}}

\bibitem[{\citenamefont{Landau and Lifshitz}(1987)}]{landau}
\bibinfo{author}{\bibfnamefont{L.~D.} \bibnamefont{Landau}} \bibnamefont{and}
  \bibinfo{author}{\bibfnamefont{E.~M.} \bibnamefont{Lifshitz}}
  (\bibinfo{year}{1987}).

\bibitem[{\citenamefont{Bhattacharyya}(2012)}]{Bhattacharyya:2012nq}
\bibinfo{author}{\bibfnamefont{S.}~\bibnamefont{Bhattacharyya}},
  \bibinfo{journal}{JHEP} \textbf{\bibinfo{volume}{1207}}, \bibinfo{pages}{104}
  (\bibinfo{year}{2012}), \eprint{1201.4654}.

\bibitem[{\citenamefont{Bhattacharyya}(2014{\natexlab{a}})}]{Bhattacharyya:2013lha}
\bibinfo{author}{\bibfnamefont{S.}~\bibnamefont{Bhattacharyya}},
  \bibinfo{journal}{JHEP} \textbf{\bibinfo{volume}{1408}}, \bibinfo{pages}{165}
  (\bibinfo{year}{2014}{\natexlab{a}}), \eprint{1312.0220}.

\bibitem[{\citenamefont{Bhattacharyya}(2014{\natexlab{b}})}]{Bhattacharyya:2014bha}
\bibinfo{author}{\bibfnamefont{S.}~\bibnamefont{Bhattacharyya}}
  (\bibinfo{year}{2014}{\natexlab{b}}), \eprint{1403.7639}.

\bibitem[{\citenamefont{Haehl et~al.}(2014)\citenamefont{Haehl, Loganayagam,
  and Rangamani}}]{Haehl:2013hoa}
\bibinfo{author}{\bibfnamefont{F.~M.} \bibnamefont{Haehl}},
  \bibinfo{author}{\bibfnamefont{R.}~\bibnamefont{Loganayagam}},
  \bibnamefont{and}
  \bibinfo{author}{\bibfnamefont{M.}~\bibnamefont{Rangamani}},
  \bibinfo{journal}{JHEP} \textbf{\bibinfo{volume}{1403}}, \bibinfo{pages}{034}
  (\bibinfo{year}{2014}), \eprint{1312.0610}.

\bibitem[{\citenamefont{Haehl et~al.}(2015{\natexlab{a}})\citenamefont{Haehl,
  Loganayagam, and Rangamani}}]{Haehl:2015pja}
\bibinfo{author}{\bibfnamefont{F.~M.} \bibnamefont{Haehl}},
  \bibinfo{author}{\bibfnamefont{R.}~\bibnamefont{Loganayagam}},
  \bibnamefont{and} \bibinfo{author}{\bibfnamefont{M.}~\bibnamefont{Rangamani}}
  (\bibinfo{year}{2015}{\natexlab{a}}), \eprint{1502.00636}.

\bibitem[{\citenamefont{Liu}(1972)}]{Liu:1972nr}
\bibinfo{author}{\bibfnamefont{I.-S.} \bibnamefont{Liu}},
  \bibinfo{journal}{Archive for Rational Mechanics and Analysis}
  \textbf{\bibinfo{volume}{46}}, \bibinfo{pages}{131} (\bibinfo{year}{1972}).

\bibitem[{\citenamefont{Loganayagam}(2011)}]{Loganayagam:2011mu}
\bibinfo{author}{\bibfnamefont{R.}~\bibnamefont{Loganayagam}}
  (\bibinfo{year}{2011}), \eprint{1106.0277}.

\bibitem[{\citenamefont{Banerjee et~al.}(2012)\citenamefont{Banerjee,
  Bhattacharya, Bhattacharyya, Jain, Minwalla et~al.}}]{Banerjee:2012iz}
\bibinfo{author}{\bibfnamefont{N.}~\bibnamefont{Banerjee}},
  \bibinfo{author}{\bibfnamefont{J.}~\bibnamefont{Bhattacharya}},
  \bibinfo{author}{\bibfnamefont{S.}~\bibnamefont{Bhattacharyya}},
  \bibinfo{author}{\bibfnamefont{S.}~\bibnamefont{Jain}},
  \bibinfo{author}{\bibfnamefont{S.}~\bibnamefont{Minwalla}},
  \bibnamefont{et~al.}, \bibinfo{journal}{JHEP}
  \textbf{\bibinfo{volume}{1209}}, \bibinfo{pages}{046} (\bibinfo{year}{2012}),
  \eprint{1203.3544}.

\bibitem[{\citenamefont{Jensen et~al.}(2012)\citenamefont{Jensen, Kaminski,
  Kovtun, Meyer, Ritz et~al.}}]{Jensen:2012jh}
\bibinfo{author}{\bibfnamefont{K.}~\bibnamefont{Jensen}},
  \bibinfo{author}{\bibfnamefont{M.}~\bibnamefont{Kaminski}},
  \bibinfo{author}{\bibfnamefont{P.}~\bibnamefont{Kovtun}},
  \bibinfo{author}{\bibfnamefont{R.}~\bibnamefont{Meyer}},
  \bibinfo{author}{\bibfnamefont{A.}~\bibnamefont{Ritz}}, \bibnamefont{et~al.},
  \bibinfo{journal}{Phys.Rev.Lett.} \textbf{\bibinfo{volume}{109}},
  \bibinfo{pages}{101601} (\bibinfo{year}{2012}), \eprint{1203.3556}.

\bibitem[{\citenamefont{Dubovsky et~al.}(2012)\citenamefont{Dubovsky, Hui,
  Nicolis, and Son}}]{Dubovsky:2011sj}
\bibinfo{author}{\bibfnamefont{S.}~\bibnamefont{Dubovsky}},
  \bibinfo{author}{\bibfnamefont{L.}~\bibnamefont{Hui}},
  \bibinfo{author}{\bibfnamefont{A.}~\bibnamefont{Nicolis}}, \bibnamefont{and}
  \bibinfo{author}{\bibfnamefont{D.~T.} \bibnamefont{Son}},
  \bibinfo{journal}{Phys.Rev.} \textbf{\bibinfo{volume}{D85}},
  \bibinfo{pages}{085029} (\bibinfo{year}{2012}), \eprint{1107.0731}.

\bibitem[{\citenamefont{Bhattacharya et~al.}(2013)\citenamefont{Bhattacharya,
  Bhattacharyya, and Rangamani}}]{Bhattacharya:2012zx}
\bibinfo{author}{\bibfnamefont{J.}~\bibnamefont{Bhattacharya}},
  \bibinfo{author}{\bibfnamefont{S.}~\bibnamefont{Bhattacharyya}},
  \bibnamefont{and}
  \bibinfo{author}{\bibfnamefont{M.}~\bibnamefont{Rangamani}},
  \bibinfo{journal}{JHEP} \textbf{\bibinfo{volume}{1302}}, \bibinfo{pages}{153}
  (\bibinfo{year}{2013}), \eprint{1211.1020}.

\bibitem[{\citenamefont{Callan and Harvey}(1985)}]{Callan:1984sa}
\bibinfo{author}{\bibfnamefont{J.}~\bibnamefont{Callan},
  \bibfnamefont{Curtis~G.}} \bibnamefont{and}
  \bibinfo{author}{\bibfnamefont{J.~A.} \bibnamefont{Harvey}},
  \bibinfo{journal}{Nucl.Phys.} \textbf{\bibinfo{volume}{B250}},
  \bibinfo{pages}{427} (\bibinfo{year}{1985}).

\bibitem[{\citenamefont{Haehl et~al.}(2015{\natexlab{b}})\citenamefont{Haehl,
  Loganayagam, and Rangamani}}]{Haehl:2015fk}
\bibinfo{author}{\bibfnamefont{F.~M.} \bibnamefont{Haehl}},
  \bibinfo{author}{\bibfnamefont{R.}~\bibnamefont{Loganayagam}},
  \bibnamefont{and}
  \bibinfo{author}{\bibfnamefont{M.}~\bibnamefont{Rangamani}},
  \bibinfo{journal}{to appear}  (\bibinfo{year}{2015}{\natexlab{b}}).

\bibitem[{\citenamefont{Baier et~al.}(2008)\citenamefont{Baier, Romatschke,
  Son, Starinets, and Stephanov}}]{Baier:2007ix}
\bibinfo{author}{\bibfnamefont{R.}~\bibnamefont{Baier}},
  \bibinfo{author}{\bibfnamefont{P.}~\bibnamefont{Romatschke}},
  \bibinfo{author}{\bibfnamefont{D.~T.} \bibnamefont{Son}},
  \bibinfo{author}{\bibfnamefont{A.~O.} \bibnamefont{Starinets}},
  \bibnamefont{and} \bibinfo{author}{\bibfnamefont{M.~A.}
  \bibnamefont{Stephanov}}, \bibinfo{journal}{JHEP}
  \textbf{\bibinfo{volume}{0804}}, \bibinfo{pages}{100} (\bibinfo{year}{2008}),
  \eprint{0712.2451}.

\bibitem[{\citenamefont{Bhattacharyya
  et~al.}(2008{\natexlab{a}})\citenamefont{Bhattacharyya, Hubeny, Minwalla, and
  Rangamani}}]{Bhattacharyya:2008jc}
\bibinfo{author}{\bibfnamefont{S.}~\bibnamefont{Bhattacharyya}},
  \bibinfo{author}{\bibfnamefont{V.~E.} \bibnamefont{Hubeny}},
  \bibinfo{author}{\bibfnamefont{S.}~\bibnamefont{Minwalla}}, \bibnamefont{and}
  \bibinfo{author}{\bibfnamefont{M.}~\bibnamefont{Rangamani}},
  \bibinfo{journal}{JHEP} \textbf{\bibinfo{volume}{0802}}, \bibinfo{pages}{045}
  (\bibinfo{year}{2008}{\natexlab{a}}), \eprint{0712.2456}.

\bibitem[{\citenamefont{Bhattacharyya
  et~al.}(2008{\natexlab{b}})\citenamefont{Bhattacharyya, Loganayagam, Mandal,
  Minwalla, and Sharma}}]{Bhattacharyya:2008mz}
\bibinfo{author}{\bibfnamefont{S.}~\bibnamefont{Bhattacharyya}},
  \bibinfo{author}{\bibfnamefont{R.}~\bibnamefont{Loganayagam}},
  \bibinfo{author}{\bibfnamefont{I.}~\bibnamefont{Mandal}},
  \bibinfo{author}{\bibfnamefont{S.}~\bibnamefont{Minwalla}}, \bibnamefont{and}
  \bibinfo{author}{\bibfnamefont{A.}~\bibnamefont{Sharma}},
  \bibinfo{journal}{JHEP} \textbf{\bibinfo{volume}{0812}}, \bibinfo{pages}{116}
  (\bibinfo{year}{2008}{\natexlab{b}}), \eprint{0809.4272}.

\bibitem[{\citenamefont{Loganayagam}(2008)}]{Loganayagam:2008is}
\bibinfo{author}{\bibfnamefont{R.}~\bibnamefont{Loganayagam}},
  \bibinfo{journal}{JHEP} \textbf{\bibinfo{volume}{0805}}, \bibinfo{pages}{087}
  (\bibinfo{year}{2008}), \eprint{0801.3701}.

\bibitem[{\citenamefont{Haack and Yarom}(2009)}]{Haack:2008xx}
\bibinfo{author}{\bibfnamefont{M.}~\bibnamefont{Haack}} \bibnamefont{and}
  \bibinfo{author}{\bibfnamefont{A.}~\bibnamefont{Yarom}},
  \bibinfo{journal}{Nucl.Phys.} \textbf{\bibinfo{volume}{B813}},
  \bibinfo{pages}{140} (\bibinfo{year}{2009}), \eprint{0811.1794}.

\bibitem[{\citenamefont{York and Moore}(2009)}]{York:2008rr}
\bibinfo{author}{\bibfnamefont{M.~A.} \bibnamefont{York}} \bibnamefont{and}
  \bibinfo{author}{\bibfnamefont{G.~D.} \bibnamefont{Moore}},
  \bibinfo{journal}{Phys.Rev.} \textbf{\bibinfo{volume}{D79}},
  \bibinfo{pages}{054011} (\bibinfo{year}{2009}), \eprint{0811.0729}.

\bibitem[{\citenamefont{Kovtun et~al.}(2005)\citenamefont{Kovtun, Son, and
  Starinets}}]{Kovtun:2004de}
\bibinfo{author}{\bibfnamefont{P.}~\bibnamefont{Kovtun}},
  \bibinfo{author}{\bibfnamefont{D.}~\bibnamefont{Son}}, \bibnamefont{and}
  \bibinfo{author}{\bibfnamefont{A.}~\bibnamefont{Starinets}},
  \bibinfo{journal}{Phys.Rev.Lett.} \textbf{\bibinfo{volume}{94}},
  \bibinfo{pages}{111601} (\bibinfo{year}{2005}), \eprint{hep-th/0405231}.

\end{thebibliography}

\end{document}